\documentclass[10pt,letterpaper,twocolumn]{article} %% two column, final layout

\usepackage{ol}
\usepackage{hyperref}
\usepackage{amsmath}
\usepackage{graphicx}

\begin{document}

\twocolumn[ %% activate for two-column option

\title{Subwavelength Diffraction Management}

\author{Matteo Conforti, Massimiliano Guasoni, and Costantino De Angelis}

\address{Consorzio Nazionale Interuniversitario per le Scienze Fisiche della Materia,
Dipartimento di Elettronica per l'Automazione, Universit\`a di Brescia, via Branze 38, 25123 Brescia, Italy}

% Do not use \email or \homepage here. E-mail and URL can be given just before references.

\begin{abstract}
We study light propagation in nanoscale periodic structures
composed of dielectric and metal in the visible range. We
demonstrate that diffraction can be tailored both in magnitude and
in sign by varying the geometric features of the waveguides.
Diffraction management on a subwavelength scale is demonstrated by
numerical solution of Maxwell equations in frequency domain.
\end{abstract}

] %% activate for two-column option

The miniaturization of photonic devices for confining and guiding
electromagnetic energy down to nanometer scale is one of the
biggest challenges for the information technology
industries\cite{maier07}. In the last years, photonic crystals
technology allowed to gain one order of magnitude factor in the
miniaturization of components such as waveguides and couplers with
respect to conventional (i.e. based on total internal reflection)
optics. However when the size of a conventional optical circuit is
reduced to the nanoscale, the propagation of light is limited by
diffraction. One way to overcome this limit is through surface
plasmon polaritons \cite{ecomomu69}, which are evanescent waves
trapped at the interface between a medium with positive dielectric
constant and one with negative dielectric constant, such as metals
in the visible range. Even though this phenomenon has been known
for a long time, in the last years there is a renewed interest in
this field, mainly motivated by the will to merge integrated
electronic circuits to photonic devices\cite{ozbay06}.

In this Letter we study the propagation of light in nanostructured
metal-dielectric waveguide arrays (plasmonic arrays). As well known,
an array of evanescently coupled single-mode waveguides exhibits
power exchanges among the waveguides leading to discrete
diffraction.\cite{Christodoulides88,Eisenberg98}. In plasmonic
arrays we find peculiar diffractive phenomena and we show the
possibility of diffraction management on a subwavelength scale.

The basic building block of a uniform waveguide array is the
directional coupler; in the following, for the sake of clarity, we
will consider only two dimensional cases: we have translational
invariance along the $z$ axis, with $y$ being the propagation
direction and $x$ the other transverse coordinate. In its simplest
form, a directional coupler consists of two identical parallel
waveguides in close proximity; as well known, the power exchange
between the two waveguides can be described by ordinary
differential equations coupling the modal field amplitude
$A_{1,2}(y)$ of waveguides 1 and 2:
\begin{eqnarray}
i \frac{d A_{1}}{d y}  + \beta \,A_{1} +C A_{2} =0 \nonumber
\\
i \frac{d A_{2}}{d y}  + \beta \,A_{2} +C A_{1} =0 \label{coupler}
\end{eqnarray}
where $\beta$ is the propagation constant of the waveguide mode
and $C$ is the coupling coefficient, whose expression can be
obtained in the framework of Coupled Mode Theory. Limiting our
attention to the TM case (i.e the non vanishing field components
are $H_z$,$E_x$,$E_y$) we have an expression valid both for
$y$-homogeneous and $y$-periodic (photonic crystal) waveguides
\cite{Hardy85, Michaelis03}:
\begin{equation}
 C = \omega \frac{\int_{-\infty}^{\infty} \int_{-L_y/2}^{L_y/2}
(\varepsilon - \varepsilon_1) \left( e_{x1}^*e_{x2}+ \displaystyle
\frac{\varepsilon_2}{\varepsilon}\,e_{y1}^*e_{y2} \right)dx dy }{2
L_y \, \Re e \left( \int_{-\infty}^{\infty} e_{x1}h_{z1}^* \, dx
\right)}, \label{cmt}
\end{equation}
where $\varepsilon(x,y)$ is the permittivity of the directional
coupler, $\varepsilon_{1,2}(x,y)$ is the permittivity of the
isolated single waveguide 1 or 2, $e_{x1,2}(x,y)$, $e_{y1,2}(x,y)$
and $h_{z1,2}(x,y)$ are the electric and magnetic field of the
mode (or the Bloch mode in case of periodic waveguide) of
waveguide 1 or 2, $L_y$ is the unit cell length for a photonic
crystal structure and an arbitrary length in the case of
translational invariance along $y$.

%In contrast with conventional optical devices based on total
%internal reflection where $C$ must be always positive, for photonic
%crystal directional couplers the coupling coefficient $C$ can be
%either positive or negative \cite{desterke04, locatelli05}.

In contrast with conventional optical devices based on total
internal reflection where $C$ must be always positive, in the
following we introduce two different plasmonic directional
couplers having opposite sign of the coupling coefficient.
%, thus
%proving that $C$ can be either positive or negative in plasmonic
%directional couplers.
\begin{figure}[tb]
\begin{center}
\includegraphics[width=4.2cm]{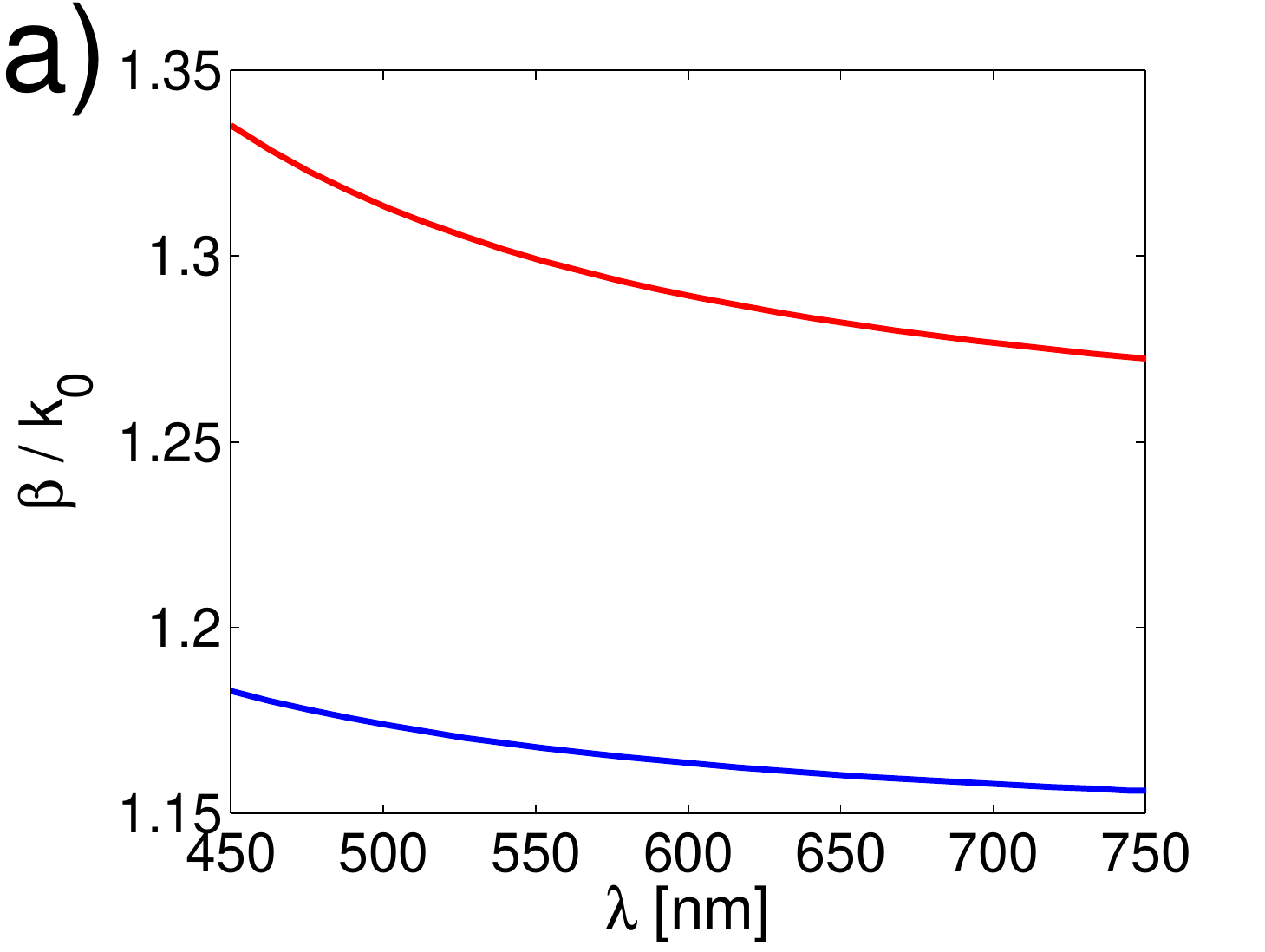}
\includegraphics[width=4.2cm]{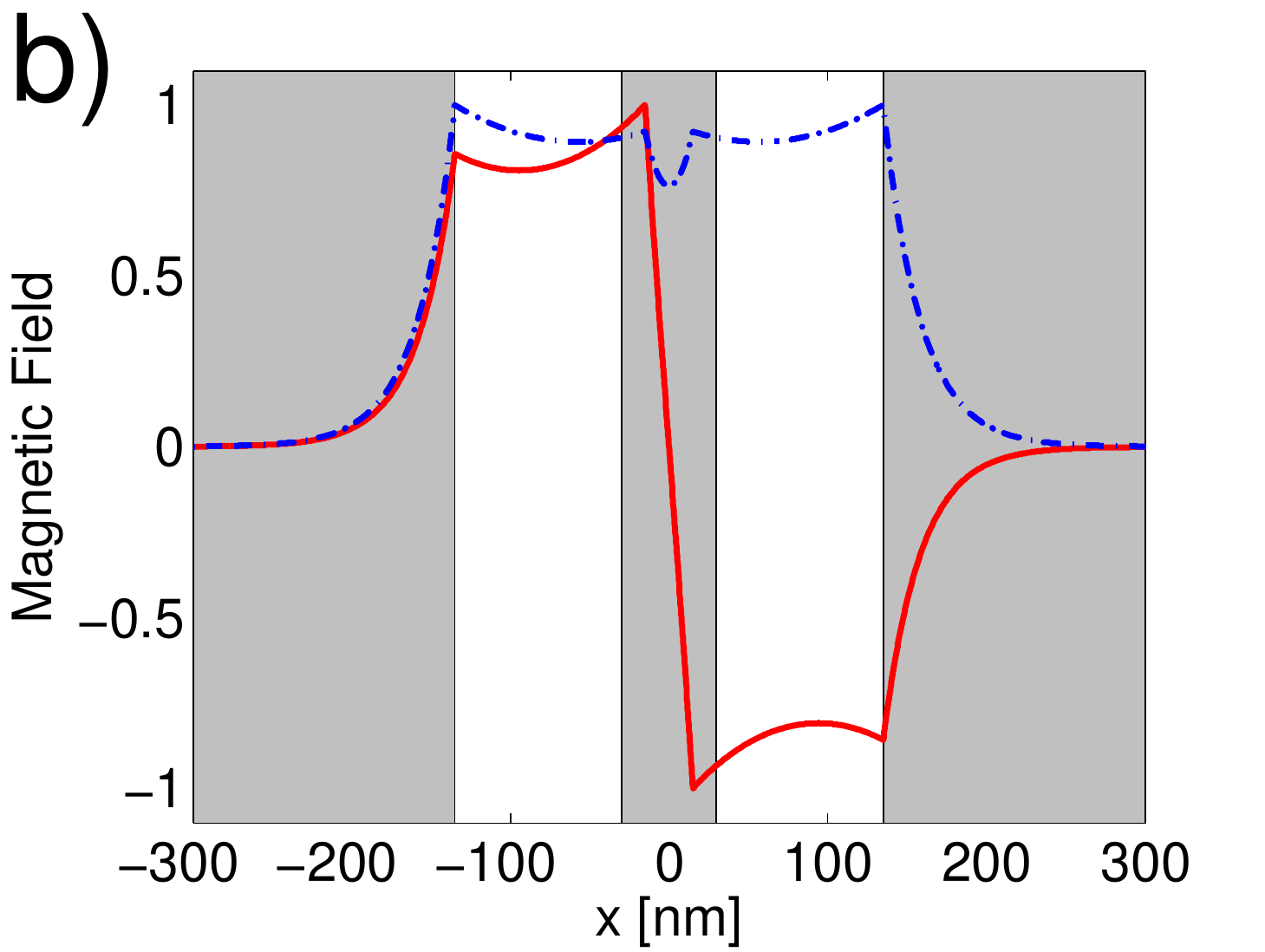}
\end{center}
 \caption{(Color online)  a) Dispersion relation of the
  fundamental (odd) and second order (even) mode of the coupler DC1.
  b) Fundamental (continuous red) and second order (dash-dot blue) mode @ 600nm. }\label{coupler1}
\end{figure}

Let us first consider a system composed of alternate layers of metal
(for example Silver) and dielectric (air). We describe the optical
properties of the metal using a Drude free-electron model
$\varepsilon(\omega)=1-\omega_p^2/[\omega(\omega-i\omega_\tau)]$,
where $\omega_p$ is the plasma frequency and $\omega_\tau$ is the
collision frequency.
%It is well known that surface plasmons at the
%interfaces between metal and dielectric can exist only for TM
%polarization.
As an example we consider a structure composed of 30nm of Silver
and 120nm of air. We neglect absorption in metal ($\omega_\tau=0$)
since it does not affect strongly the dispersive properties of the
propagating modes. In Fig. \ref{coupler1} we show the dispersion
relation and the profiles of the two modes supported by this
directional coupler (to be called DC1 in the rest of the paper).
In stark contrast with conventional waveguides, the fundamental
mode of this structure is odd\cite{fan06}. Moreover we can see
that the fundamental mode has one node whereas the second one has
no. This feature seems to struggle against the well known Sturm
oscillation theorem\cite{zettl05}, which states that the $N$th
order mode has exactly $N-1$ nodes ($N=1,2,\ldots$); however it
has to be remembered that the theorem holds only if the dielectric
constant is everywhere positive. \noindent Note also that the
reversal of modes parity implies a negative value of the coupling
coefficient\cite{desterke04,locatelli05} $C=(\beta_{even} -
\beta_{odd})/2$. For example at a wavelength of 600 nm
$C_{\Delta\beta}=(\beta_{even} - \beta_{odd})/2=-6.59\cdot
10^{5}m^{-1}$ and from Eq. (\ref{cmt}) $C_{CMT}=-5.75\cdot 10^5
m^{-1}$. The difference between the two values (around 13\%) is
due to the strong coupling between the waveguides.

%, leading to an anomalous
%diffraction coefficient at normal incidence ($k_x=0$).

\begin{figure}[tb]
\begin{center}
\includegraphics[width=4.4cm]{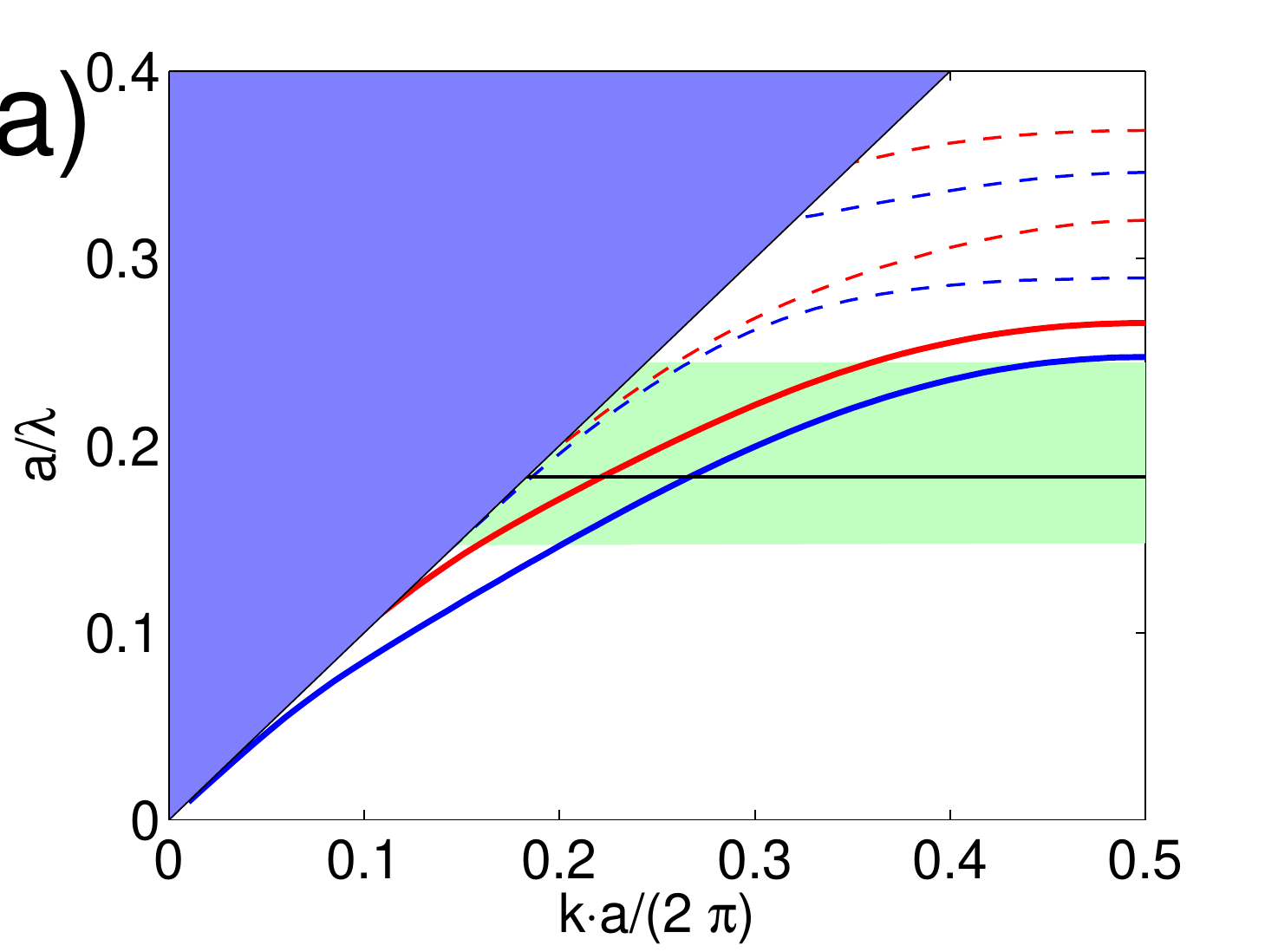}
\includegraphics[width=4cm]{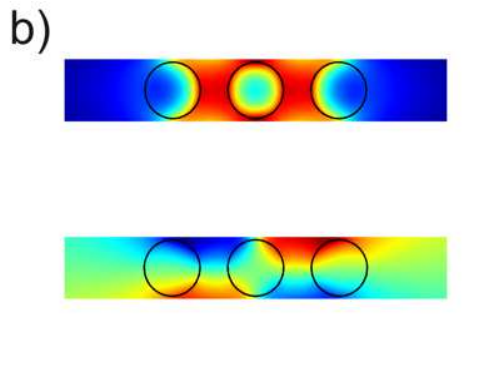}
\end{center}
 \caption{(Color online)  a) Dispersion relation of the coupler DC2.
 Thick curves, fundamental (even) and second order (odd) mode. Dashed curves, high order modes.
 Thin horizontal line denotes a wavelength of 600 nm. Dark shaded region, light cone; light shaded region, operating bandwidth
 450nm-750nm.
   b) Fundamental and second order mode @ 600nm. }\label{coupler2}
\end{figure}

Another well known plasmonic guiding structure is the metal
nanoparticle array\cite{quinten98}, where the energy transport is
supplied by electromagnetic resonant coupling between metal
particles arranged in a linear chain. Double nanoparticle chains,
where the electromagnetic energy is confined between two linear
chains, offer a more flexible structure since the propagation is
less determined by resonances\cite{chu07}, allowing for a larger
bandwidth for the guided modes. As an example we considered double
chain waveguides composed of Silver nanoparticles with a radius of
50 nm immersed in air with a longitudinal separation of 110 nm and
a distance between the chains of 150 nm. We neglected losses in
the metal since it was found from previous band diagram
calculations of metallic photonic crystals
%, in which a plasmonic model was used,
that this is reasonable for realistic absorption coefficients
\cite{catrysse06}. In Fig. \ref{coupler2} we show the dispersion
relation and the mode profiles of two coupled waveguides composed
of three nanoparticle chains (to be called directional coupler DC2
in the rest of the paper). Since we are interested in guided modes
we consider only the region in the $(k,\omega)$ space that lies
below the light line $\omega=ck$. This coupler supports several
modes, however in the working range [450nm-750nm] (light shaded
region in Fig. \ref{coupler2}a)) only two modes are reasonably
below the light line. As opposed to the previous directional
coupler, the fundamental mode here is even and the second order
mode is odd, implying positive coupling constant. For example at
600nm we obtain $C_{\Delta\beta}=1.29\cdot 10^{6}m^{-1}$ and
$C_{CMT}=1.15\cdot 10^6 m^{-1}$.

Let us now consider the uniform arrays A1 (A2) obtained using
directional couplers DC1 (DC2) as basic building blocks. The set
of equations modelling the evolution of the modal field amplitude
$A_n$ in each waveguide of an uniform array reads
as:\cite{Christodoulides88,Eisenberg98}
\begin{equation}
i \frac{d A_{n}}{d y}  + \beta \,A_{n} +C (A_{n+1}+A_{n-1}) =0.
\label{set_2}
\end{equation}
The diffraction coefficient\cite{Eisenberg00,Pertsch02} of the
waveguide array can be derived from Eqs. (\ref{set_2}):
\begin{equation}
D= -2 C d^2 \cos(k_x d) \label{diffraction},
\end{equation}
where $d$ is the spacing between the centers of adjacent waveguides
and then $k_x d $ is the imposed input field phase shift between
them.

From the above reported analysis we expect, at normal incidence
($k_x=0$), the waveguide array A1 to be an anomalous-diffraction
array ($D>0$), whereas waveguide array A2 is expected to behave as
a normal-diffraction array ($D<0$). The field evolution along the
waveguide arrays has been simulated without any approximation by
solving Maxwell's equations through a frequency-domain
finite-element method, using arrays composed by 17 waveguides. The
central waveguide of the arrays is excited with a Gaussian field,
which spreads during propagation and generates the typical
diffraction pattern observed also in conventional waveguide
arrays: two outermost wings and a few less intense peaks in the
central waveguides. The same qualitative behavior is observed for
both arrays since the intensity evolution is not influenced by the
diffraction sign. On the other hand the phase front curvature of
the propagating field depends on the diffraction sign. Therefore,
if we alternate arrays characterized by normal and anomalous
diffraction, the input field shape can be periodically recovered.

\begin{figure}[tb]
\begin{center}
\includegraphics[width=7cm]{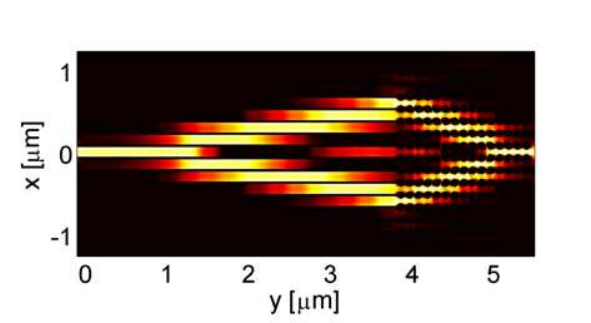}
\end{center}
 \caption{(Color online) Time average power flow in $y$ direction
  (normalized to the maximum) in the diffraction managed device @ 600nm. }\label{3}
\end{figure}

In Fig. \ref{3}  we report the finite element simulation of the
diffraction-managed device. The first array section (A1) is
$L_1$=3900nm long, the second (A2) $L_2$=1600nm. We excited only
the central waveguide with a normal-incident 120nm FHWM intensity
Gaussian field at 600nm. The input excitation spreads during
propagation in the first section, whereas it exhibits an opposite
behavior in the second one, and the initial field distribution is
recovered at the output end of the device. The reported phenomena
are a clear signature of the inversion of the sign of the
diffraction coefficient, in perfect agreement with our theoretical
analysis. It is worth noting that all the dynamics takes place in
 a $2\mu m\times 5\mu m$ device.
%This size is much lower than other similar
%photonic-crystal based devices proposed before.

%Since we are dealing with metals a few words has to be spent for
%the discussion of losses.
As far as losses are concerned, we verified that the propagation
in the diffraction managed devices is not influenced at all by
including a lossy model for the metal (Drude model gives
$\varepsilon(600nm)\approx -15 - 0.37i $): the evolution showed in
Fig. \ref{3} is indistinguishable from the analogue calculated
with a lossy metal. Moreover the decay length of the fundamental
mode of the waveguides composing A1 is, for example,
$L_D(600nm)=[2\rm{Im}(\beta)]^{-1}= 17\mu m$, much longer than the
device length, indicating that all the relevant dynamics can take
place without being suppressed by absorption.

\begin{figure}[tb]
\begin{center}
\includegraphics[width=4.2cm]{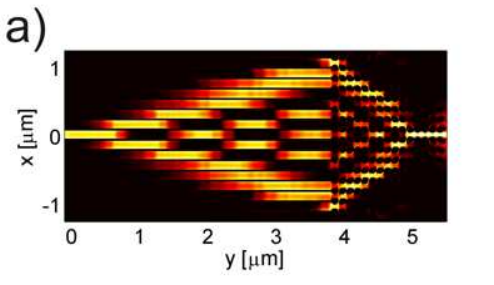}
\includegraphics[width=4.2cm]{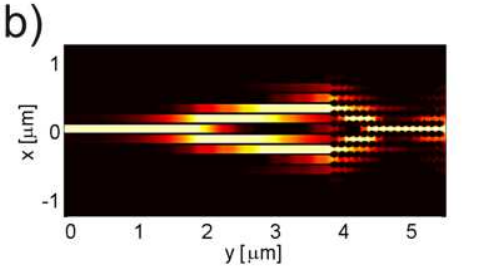}\\
\includegraphics[width=4.2cm]{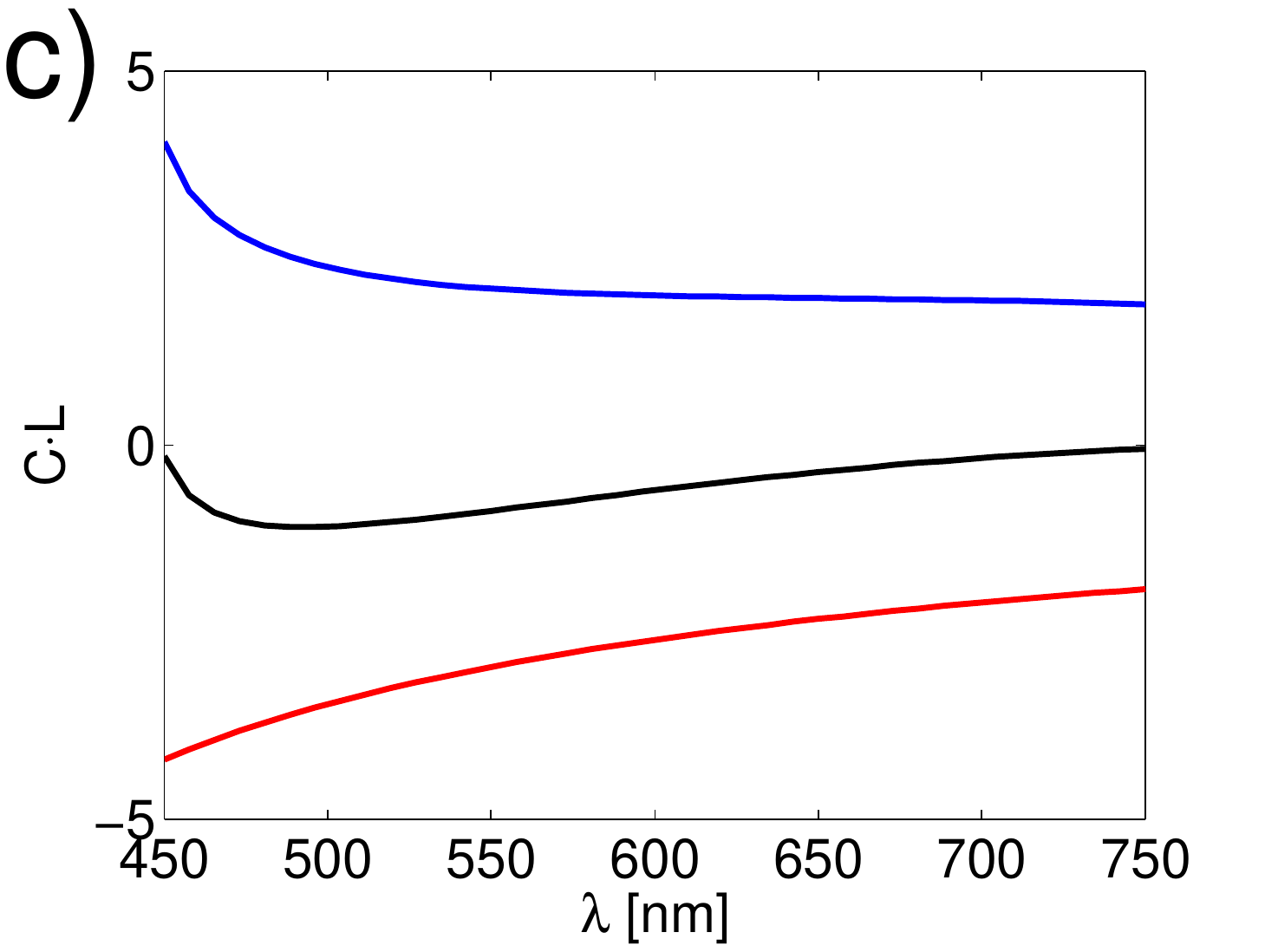}

\end{center}
 \caption{(Color online)  Time average power flow  in $y$ direction (normalized to the maximum)
 in the diffraction managed device at different
 wavelengths
 : a) 450nm, b) 750nm.
 c) Diffraction coefficient times propagation length ($C \cdot L$) for array A1 (red), array A2 (blue) and for the entire device
  ($C_1\cdot L_1 + C_2\cdot L_2$) (black). }\label{4}
\end{figure}

We now turn our attention to the bandwidth of the reported
phenomena. By observing the band structure of the metal
nanoparticle coupler (Fig. \ref{coupler2} a)), we expect the
bandwidth of this device to be the interval [450nm,750nm], i.e all
the visible range from blue to red. We studied the dynamics of
diffraction management in this range and we found that, even if
the two guiding structures have very different dispersive
properties, the diffraction compensation is reasonably good for
all the wavelengths range. Figures \ref{4}a)-b) show the time
average power flow evolution at the edges of the operating
frequencies interval. At short wavelengths (Fig. \ref{4}a)) the
magnitude of the dielectric constant of metal is relatively low
($\approx-8$), the fundamental mode of the waveguides is poorly
confined and the coupling is strong, leading to large diffraction
in both arrays. Whereas at long wavelengths (Fig. \ref{4}b)) the
large magnitude of the dielectric constant of metal ($\approx-24$)
leads to strong confinement and small diffraction. The average
value of diffractive spreading however remains low, considering
the huge band we are looking at. Figure \ref{4} c) displays the
dispersion of the diffraction parameter (coupling coefficient
times propagation length $C\cdot L$) for the arrays A1 and A2 and
for the entire device.

In conclusion, diffraction properties  of uniform arrays of
plasmonic waveguides have been studied. Starting from the analysis
of coupling between adjacent waveguides we have demonstrated that
diffraction can be controlled both in amplitude and sign with
normal incidence input excitations. Diffraction management in an
ultracompact  device composed of alternated sections with opposite
diffraction sign has been demonstrated on a wavelength interval
covering the visible range from blue to red.

\end{document}